\documentclass[conference, a4paper]{IEEEtran}
\IEEEoverridecommandlockouts

\usepackage{amsmath,amssymb,amsfonts}
\usepackage{algorithmic}
\usepackage{graphicx}
\usepackage{dblfloatfix}
\usepackage{textcomp}
\usepackage{xcolor}
\usepackage{float}
\usepackage{multirow}
\usepackage{footnote}
\usepackage{comment}
\usepackage{placeins}
\usepackage{siunitx}

\usepackage{subcaption}
\usepackage[a4paper, total={170mm, 257mm}, left=20mm, top=20mm]{geometry}

\usepackage{acronym}
\newacro{5g}[5G]{Fifth Generation}
\newacro{6g}[6G]{Sixth Generation}
\newacro{3gpp}[3GPP]{3rd Generation Partnership Project}
\newacro{as}[AS]{Authentication Server}
\newacro{atm}[ATM]{Automated Teller Machine}
\newacro{ai}[AI]{Artificial Intelligence}
\newacroplural{ais}[AIs]{Artificial Intelligences}
\newacro{ar}[AR]{autoregressive}
\newacro{b5g}[B5G]{Beyond-5G}
\newacro{ban}[BAN]{Body Area Network}
\newacro{bsi}[\textit{BSI}]{\textit{Federal Office for Information Security}}
\newacro{bdr}[BDR]{Bit Disagreement Rate}
\newacro{bs}[BS]{Base Station}
\newacro{ble}[BLE]{Bluetooth Low Energy}
\newacro{ber}[BER]{Bit Error Rate}

\newacro{ca}[CA]{Certification Authority}
\newacro{cav}[CAV]{Connected Autonomous Vehicles}
\newacro{cc}[CC]{Common Criteria}
\newacro{cir}[CIR]{Channel Impulse Response}
\newacro{cr}[CR]{Challenge-Response}
\newacro{cpu}[CPU]{Central Processing Unit}
\newacro{cpps}[CPPS]{Cyber-Physical Production System}
\newacro{crl}[CRL]{Certificate Revocation List}
\newacro{csi}[CSI]{Channel State Information}
\newacro{crke}[CRKE]{Channel-Reciprocity Based Key Extraction}
\newacro{ctf}[CTF]{Channel Transfer Function}
\newacro{cotf}[COTF]{Commercial-off-the-Shelf}
\newacro{cmos}[CMOS]{Complementary Metal-Oxide-Semiconductors}
\newacro{crlb}[CRLB]{Cramér–Rao lower bound}
\newacro{dds}[DDS]{Data Distribution Service}
\newacro{durllc}[dURLLC]{decentralized Ultra-Reliable Low Latency Communication}
\newacro{dos}[DoS]{Denial-of-Service}
\newacro{dt}[DT]{Digital Twin}
\newacro{ddos}[DDoS]{Distributed-Denial-of-Service}
\newacro{dna}[DNA]{Deoxyribonucleic Acid}
\newacro{dtls}[DTLS]{Datagram Transport Layer Security}
\newacro{dct}[DCT]{Discrete Cosine Transformation}
\newacro{dlt}[DLT]{Distributed Ledger Technology}
\newacro{dp}[DP]{Differential Privacy}
\newacro{eal}[EAL]{Evaluation Assurance Level}
\newacro{ecc}[ECC]{Elliptic Curve Cryptography}
\newacro{ecg}[ECG]{Electrocardiogram}
\newacro{eeg}[EEG]{Electroencephalogram}
\newacro{embb}[eMBB]{enhanced Mobile broad-Band}
\newacro{emg}[EMG]{Electromyogram}
\newacro{eog}[EOG]{Electrooculography}
\newacro{enb}[eNodeB]{Evolved Node B}
\newacro{er}[ER]{Extended Reality}
\newacro{etsi}[ETSI]{European Telecommunications Standards Institute}
\newacro{fpga}[FPGA]{Field Programmable Gate Array}
\newacro{fdd}[FDD]{Frequency Division Duplexing}
\newacro{fl}[FL]{Federated Learning}
\newacro{gdpr}[GDPR]{General Data Protection Regulation}
\newacro{gd}[G\&D]{Giesecke \& Devrient}
\newacro{gan}[GAN]{Generative Adversarial Network}
\newacro{h2m}[H2M]{Human-to-Machine}
\newacro{h2s}[H2S]{Human-to-Service}
\newacro{hmac}[HMAC]{Keyed-Hash Message Authentication Code}
\newacro{htc}[HTC]{Hologaphic-Type Communication}
\newacro{hotp}[HOTP]{HMAC-based One-time Password Algorithm}
\newacro{hsm}[HSM]{Hardware Security Module}
\newacro{he}[HE]{Homomorphic Encryption}
\newacro{ics}[ICS]{Industrial Control System}
\newacro{iacs}[IACS]{Industrial Automation and Control System}
\newacro{isac}[ISAC]{Integrated Sensing and Communication}
\newacro{ioe}[IoE]{Internet of Everything}
\newacro{iiot}[IIoT]{Industrial Internet of Things}
\newacro{iot}[IoT]{Internet of Things}
\newacro{io}[I/O]{Input/Output}
\newacro{ic}[IC]{Integrated Circuit}
\newacro{id}[ID]{Identificator}
\newacro{ids}[IDS]{Intursion Detection System}
\newacro{irs}[IRS]{Intelligent Reflecting Surface}
\newacro{istn}[ISTN]{Integrated Space and Terrestrial Network}
\newacro{it}[IT]{Information Technology}
\newacro{itu}[ITU]{International Telecommunication Union}
\newacro{ieee}[IEEE]{Institute of Electrical and Electronics Engineers}
\newacro{ietf}[IETF]{Internet Engineering Task Force}
\newacro{jcop}[JCOP]{Java Card Open Platform}
\newacro{jcas}[JCAS]{Joint Communication and Sensing}
\newacro{kba}[KBA]{Knowledge Based Authentication}
\newacro{kdf}[KDF]{Key Derivation Function}
\newacro{kmr}[KMR]{key mismatch rate}
\newacro{led}[LED]{Light Emitting  Diode}
\newacro{lte}[LTE]{Long Term Evolution}
\newacro{ltea}[LTE-A]{Long Term Evolution Advanced}
\newacro{lr}[LR]{Linear Regression}
\newacro{los}[LoS]{Line of Sight}
\newacro{lorawan}[LoRaWAN]{Long Range Wide Area Network}
\newacro{mbb}[MBB]{Mobile Broadband}
\newacro{mfa}[MFA]{Multi-Factor Authentication}
\newacro{mcc}[MCC]{Mobile Cloud Computing}
\newacroplural{mcus}[MCUs]{Microcontroler Units}
\newacro{m2m}[M2M]{Machine-to-Machine}
\newacro{m2s}[M2S]{Machine-to-Service}
\newacro{mimo}[MIMO]{Multiple Input Multiple Output}
\newacro{mmimo}[mMIMO]{massive Multiple Input Multiple Output}
\newacro{ml}[ML]{Machine Learning}
\newacro{mulc}[mULC]{massive Ultra-Reliable Low-Latency Communication}
\newacro{mmtc}[MMTC]{massive Machine Type Communication}
\newacro{mmg}[MMG]{Mechanomyogram}
\newacro{multos}[MULTOS]{Multii-Application Smart Card Operating System}
\newacro{mux}[MUX]{Multiplexer}
\newacro{mnc}[MNC]{Mobile Network Code}
\newacro{me}[ME]{Mobile Environment}
\newacro{mac}[MACs]{Message Authentication Codes}
\newacro{macl}[MAC]{Medium Access Control} 
\newacro{miot}[MIoT]{Medical Internet of Things}
\newacro{ngmn}[NGMN]{Next Generation Mobile Network}
\newacro{nic}[NIC]{Network Interface Controller}
\newacro{nist}[NIST]{National Institute of Standards and Technology}
\newacro{ntn}[NTN]{Non-Terrestrial Networks}
\newacro{nr}[NR]{New Radio}
\newacro{nlos}[non-Lone of Sight]{NLoS}
\newacro{oath}[OATH]{Open Authentication}
\newacro{ocra}[OCRA]{\ac{oath} Challenge-Response Algorithm}
\newacro{ocsp}[OCSP]{Online Certificate Status Protocol}
\newacro{otp}[OTP]{One-Time Password}
\newacro{oran}[O\mbox{-}RAN]{Open RAN Alliance}
\newacro{pap}[PAP]{Password-Authentication-Protocol}
\newacro{physec}[PhySec]{physical layer security}
\newacro{plkg}[PLKG]{physical layer key generation}
\newacro{pfs}[PFS]{Perfect Forward Secrecy}
\newacro{pin}[PIN]{Personal Identification Number}
\newacro{pkc}[PKC]{Public Key Cryptography}
\newacro{pki}[PKI]{Public Key Infrastructure}
\newacro{ppg}[PPG]{Photoplethysmography}
\newacro{prng}[PRNG]{Pseudo Random Number Generator}
\newacro{puf}[PUF]{Physically Unclonable Function}
\newacroplural{pufs}[PUFs]{Physically Unclonable Functions}
\newacro{pla}[PLA]{Physical Layer Authentication}
\newacro{phy}[PHY]{Physical Layer}
\newacro{pls}[PLS]{Physical Layer Security}
\newacro{plkg}[PLKG]{Physical Layer Key Generation}
\newacro{qr}[QR]{Quick Response}
\newacro{rat}[RAT]{Radio Access Technology}
\newacro{radius}[RADIUS]{Remote Authentication Dial-In User Service}
\newacro{ram}[RAM]{Random-Access Memory}
\newacro{ran}[RAN]{Radio Access Networks}
\newacro{rf}[RF]{Radio-Frequency}
\newacro{rfid}[RFID]{Radio-Frequency Identification}
\newacro{ris}[RIS]{Reconfigurable Intelligent Surface}
\newacro{rng}[RNG]{Random Number Generator}
\newacro{ro}[RO]{Ring-Oscillator}
\newacro{rom}[ROM]{Read-Only Memory}
\newacro{ros2}[ROS~2]{Robot Operating System~2}
\newacro{rs}[RS]{Reed-Solomon}
\newacro{rsa}[RSA]{Rivest-Shamir-Adleman}
\newacro{rssi}[RSSI]{Received Signal Strength Indicator}
\newacro{rsrp}[RSRP]{Reference Signal Received Power}
\newacro{re}[RE]{Resource Elements}
\newacro{rtps}[RTPS]{Real-Time Publish-Subscribe}
\newacro{ric}[RIC]{RAN Intelligent Controller}
\newacro{sdn}[SDN]{Software-Defined Network}
\newacro{sdr}[SDR]{Software-Defined Radio}
\newacro{seccos}[SECCOS]{Secure Chip Card Operating System}
\newacro{sip}[SIP]{Session Initiation Protocol}
\newacro{skg}[SKG]{Secret Key Generation}
\newacro{sram}[SRAM]{Static Random Access Memory}
\newacro{srs}[SRS]{Software Radio Systems}
\newacro{starcos}[STARCOS]{Smart Card Chip Operating System}
\newacro{sha}[SHA]{Secure Hash Algorithm}
\newacro{se}[SE]{Static Environment}
\newacro{svm}[SVM]{Support Vector Machine}
\newacro{sla}[SLA]{Service Level Agreement}
\newacro{snr}[SNR]{Signal-to-Noise-Ratio}
\newacro{tcg}[TCG]{Trusted Computing Group}
\newacro{tpm}[TPM]{Trusted Platform Module}
\newacro{tls}[TLS]{Transport Layer Security}
\newacro{trng}[TRNG]{True Random Number Generator}
\newacro{tsn}[TSN]{Time-Sensitve Networking}
\newacro{tofu}[TOFU]{Trust On First Use}
\newacro{tufu}[TUFU]{Trust Upon First Use}
\newacro{totp}[TOTP]{Time-based One-time Password Algorithm}
\newacro{tee}[TEE]{Trusted Execution Environment}
\newacro{uav}[UAV]{Unmanned Arial Vehicles}
\newacro{usb}[USB]{Universal Serial Bus}
\newacro{usrp}[USRP]{Universal Software Radio Peripheral}
\newacro{uhd}[UHD]{USRP Hardware Driver}
\newacro{usim}[USIM]{Universal Subscriber Identity Module}
\newacro{ue}[UE]{User Equipment}
\newacro{urllc}[URLLC]{Ultra-Reliable Low-Latency Communication}
\newacro{ulbc}[ULBC]{Ultra-Reliable Low-Latency Broadband Communication}
\newacro{umbb}[uMBB]{ubiquious Mobile Broadband}
\newacro{ummimo}[UM-MIMO]{Ultra-Massive MIMO}
\newacro{urdf}[URDF]{Unified Robot Description Format}
\newacro{uml}[UML]{Unified Modeling Language}
\newacro{vlc}[VLC]{Visible Light Communication}
\newacro{wban}[WBAN]{Wireless Body Area Network}
\newacro{warp}[WARP]{Wireless open-Access Research Platform}

\usepackage[
	pdfborder={0 0 0},
	colorlinks=false,%
	urlcolor=black,%
	linkcolor=black,%
	citecolor=black,%
	filecolor=black,%
	breaklinks,%
	]{hyperref}
	
\usepackage[
	backend=biber,
	style=ieee,
	isbn=false,%
	hyperref=true,%
	maxbibnames=99,%
	sorting=none,%
	natbib=true,%
	language=english,%
	]{biblatex}%
    
\DeclareFieldFormat{sentencecase}{\csname bbx@colon@search\endcsname#1}

\def\BibTeX{{\rm B\kern-.05em{\sc i\kern-.025em b}\kern-.08em
    T\kern-.1667em\lower.7ex\hbox{E}\kern-.125emX}}

\def\nobreakbefore{%
  \relax\ifvmode\else
    \ifhmode
      \ifdim\lastskip > 0pt\relax
        \unskip\nobreakspace
      \fi
    \fi
  \fi
}
\let\oldcite\cite

\renewcommand\cite[1]{\nobreakbefore{\mbox{\oldcite{#1}}}} 

\usepackage{balance} 

\makeatletter
\renewcommand\paragraph{\@startsection{paragraph}{4}{\z@}%
  {1.5ex \@plus0.5ex \@minus.2ex}
  {1ex \@plus0.2ex}
  {\normalfont\normalsize\bfseries}}
\makeatother

\begin{document}

\title{Multi-Objective RIS Deployment Optimization for Physical Layer Security in ISAC Networks 
}

\author{
\IEEEauthorblockN{Wenqing Dai\IEEEauthorrefmark{2}, Jan Herbst\IEEEauthorrefmark{2},  Jan Petershans\IEEEauthorrefmark{2}, Christoph Lipps\IEEEauthorrefmark{2}, and Hans D. Schotten\IEEEauthorrefmark{1}\IEEEauthorrefmark{2}}
\IEEEauthorblockA{\IEEEauthorrefmark{2}Intelligent Networks Research
    Group, German Research Center for Artificial Intelligence\\ D-67663
    Kaiserslautern, Email: \{firstname.lastname\}@dfki.de}
\IEEEauthorblockA{\IEEEauthorrefmark{1}Institute for Wireless
    Communication and Navigation, RPTU University Kaiserslautern-Landau\\ D-67663
    Kaiserslautern, mail: schotten@rptu.de}

}

%
%
%
%
\twocolumn[
\begin{@twocolumnfalse}
\maketitle
\begin{abstract}

Reconfigurable Intelligent Surfaces (RIS) have emerged as a key enabler for programmable wireless environments in future \ac{b5g} and 6G networks. In the meantime, Integrated Sensing and Communication (ISAC) and Physical-Layer Security (PLS) are becoming essential functionalities for next-generation wireless systems, particularly in safety and mission-critical applications. However, jointly optimizing RIS-assisted systems to support communication, sensing, and security introduces complex and often conflicting design trade-offs.
In this work, a multi-objective optimization framework for RIS-assisted networks is proposed, aiming to jointly analyze communication performance, sensing accuracy, and security-related channel properties in a unified system perspective. The proposed model jointly considers RIS deployment location, orientation, surface size, and an ISAC configuration weight that controls the allocation of RIS reflection gain between communication and sensing tasks.
Simulation results reveal inherent trade-offs among communication reliability, sensing accuracy, and security performance. The proposed framework provides valuable insights into the interplay between communication, sensing, and security, and enables the design of efficient RIS deployment and configuration strategies for secure ISAC-enabled 6G wireless networks.\newline

\noindent\emph{Please note: This is a preprint of a paper which has been accepted for publication at Mobile Communication - Technologies and Applications; 30th ITG Symposium 2026.}

\end{abstract}

\end{@twocolumnfalse}
]

%
%
%
%

\label{sec:introduction}

\section{Introduction}
Future \ac{b5g} and \ac{6g} wireless networks will require not only high-speed data transmission, but also environmental sensing. \ac{isac} is regarded as a key technological direction for next-generation wireless systems. By sharing hardware and spectrum resources, \ac{isac} enables communication systems to perform environmental sensing while transmitting data, thereby improving spectrum utilization efficiency and expanding network functionality \cite{11111722}. 
Beyond these functionalities, future systems are increasingly required to exhibit verifiable and trustworthy behavior, particularly in safety-critical and mission-critical applications. Thus, inherent security capabilities are becoming increasingly relevant to trustworthy \ac{6g}. 

At the same time, as an emerging technology, \acp{ris} offer new solutions for creating programmable wireless propagation environments \cite{Renzo.2020}. \acp{ris} comprises a large number of tunable reflective elements, each capable of dynamically adjusting the reflection of incident electromagnetic waves \cite{Wu.2021}. By appropriately configuring these reflective elements, a \ac{ris} can effectively reshape wireless propagation channels. This can enhance communication coverage, reducing interference, and improving overall performance of the wireless communication system, for example in terms of signal quality, reliability, and achievable data rates.
Beyond these improvements, \acp{ris} enable a controlled modification of propagation conditions, which directly affects the structure of channel measurements used for sensing, localization, and security-related inference. Recent research has shown that \acp{ris} can enhance legitimate links by modulating channel propagation characteristics and by shaping the channel-dependent features exploited by \ac{pls} mechanisms such as key generation, device authentication, and spoofing detection, thereby improving \ac{pls} performance \cite{Lipps.2024, dai2025itg}.

However, simultaneously optimizing communication, sensing, and security performance in \ac{ris}-assisted \ac{isac} networks remains a challenge. \ac{ris} configurations optimized for communication performance may fail to provide sufficient sensing coverage, while configurations optimized for sensing accuracy may inadvertently enhance signal strength in the direction of eavesdroppers \cite{11373496, 10979924,Chopra2025RISISAC}. Furthermore, channel characteristics favorable for sensing do not necessarily improve the channel reciprocity required for key generation \cite{10989512, dai2025itg}. 

This work proposes a multi-objective optimization framework for \ac{ris}-assisted networks, aiming to jointly analyze communication performance, sensing accuracy, and security-related channel properties in a unified system perspective. The \ac{ris} configuration is modeled as a multi-objective optimization problem, with objectives including maximizing the communication \ac{snr}, the security gap between the legitimate and eavesdropping links, and the \ac{ris}-enabled sensing gain over the direct path. The key contributions of this work are summarized as follows:

\begin{itemize}
    \item Proposal of \ac{ris} deployment in a secure \ac{isac} system, with a detailed model capturing channel dynamics.
    \item Development of a unified \ac{ris} system model that jointly considers communication performance, sensing accuracy, and communication security.
    \item Comprehensive performance analysis based on the key quality metrics \ac{snr}, sensing gain at target end, and security gap between legitimate user and eavesdropper, providing insights into favorable \ac{ris} deployment scenarios.
\end{itemize}

The remainder of this work is structured as follows.
To provide an overview, \autoref{state-of-the-art} highlights the current research activities. The system topology is introduced, and channel modeling is conducted in \autoref{experimental-setup}. \autoref{result} presents the simulation setup and performance evaluation, followed by \autoref{conclusion}, concluding the study and discussing future research directions.

\section{Related Work} 
\label{state-of-the-art}

\ac{ris} have recently emerged as a key enabler for programmable wireless environments by intelligently manipulating electromagnetic wave propagation. In the following, a brief overview of this topic and related research is outlined.

\subsection{RIS-enabled Smart Radio Environments and ISAC Systems}
The concept of smart radio environments was established by \textit{Di Renzo et al.}, who demonstrated how \ac{ris} can transform wireless channels into controllable entities and outlined future research directions \cite{Renzo.2020}. Subsequent surveys further summarize the rapid advances of \ac{ris} architectures, signal processing, and optimization, including emerging extensions such as STAR-\ac{ris} and \ac{ai}-assisted control for millimeter-wave and THz communication in future 6G systems  \cite{Hassouna2023ASO, Iqbal2025ACS}.

More recently, \acp{ris} have been increasingly studied in \ac{isac} systems, where controllable reflections jointly support communication and localization \cite{Chopra2025RISISAC, Mai.2025}. In this context, \acp{ris} have been shown to improve channel observability and localization accuracy, for example in indoor positioning \cite{Ma.2021}. Likewise, \textit{Liu et al.} analyze multiple-\ac{ris}-assisted millimeter-wave positioning systems, deriving the \ac{crlb} to quantify localization accuracy under different configurations \cite{Liu2021OptimizationOR}. Recent studies further demonstrate that \ac{crlb}-driven phase optimization can effectively reduce localization error, particularly in near-field scenarios \cite{Macias.2024, Shaikh.2024}. Overall, \ac{ris}-enabled \ac{isac} systems offer considerable potential, but also involve non-trivial trade-offs across communication and sensing objectives.

\subsection{RIS-assisted PLS}
In addition to sensing, \ac{ris} has also received growing attention in \ac{pls} and secret key generation. Prior work shows that intelligent reflection control can enhance secrecy performance by strengthening legitimate channels while suppressing eavesdropping links \cite{Lipps.2024}. Meanwhile, \ac{plkg} has emerged as a lightweight security mechanism that exploits channel reciprocity and randomness to establish secret keys \cite{9657177}, offering advantages over conventional cryptographic methods in next-generation networks \cite{Xiao.2024}. For instance, \textit{Lu et al.} propose an \ac{ris}-assisted probing protocol combined with Grassmannian-based optimization to improve key generation performance in low-entropy environments \cite{Lu.2023}, while experimental validation of \ac{ris}-enhanced channel randomness using OFDM systems is provided in \cite{Staat.2021}. Considered together, the existing literature indicates that \ac{ris} not only benefits communication and sensing performance, but also contributes significantly to improved wireless security \cite{lipps2023ris,9984880}.

\section{Experimental setup}
\label{experimental-setup}

This section presents the experimental setup and methodology employed in this work. It first introduces the topology of the considered scenario, followed by the mathematical models and performance metrics used for the analysis. Finally, the underlying optimization objective is formulated.

\subsection{Scenario and Topology}

\begin{figure}[b]
    \centering
    \includegraphics[width=0.5\textwidth]{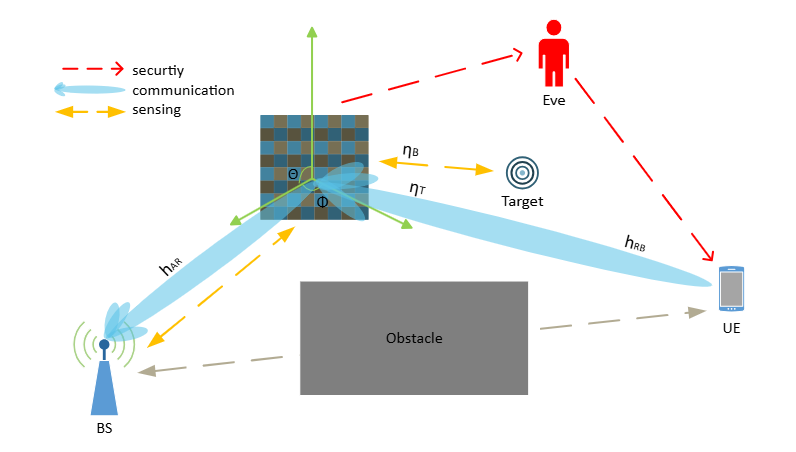}
    \caption{Schematic diagram of the system topology.}
    \label{fig:topology} 
\end{figure}

A \ac{ris}-assisted wireless network is considered, designed to support integrated sensing, communication, and security functionalities. A schematic overview of the system is outlined in \autoref{fig:topology}, comprising five main entities: a base station (BS), a legitimate receiver (Bob), a potential eavesdropper (Eve), a sensing target, and a deployable \ac{ris}. The BS acts as the transmitter and performs both communication and sensing tasks. Bob represents the legitimate communication user communicating with the BS. Eve attempts to intercept the transmitted signal. Meanwhile, the target node is used for sensing purposes. 

The positions of BS, Bob, Eve, and the sensing target are fixed, while the \ac{ris} location can be optimized within the deployment area of 100 $\times$ 100 m, as depicted in \autoref{fig:heatmap}. This specific layout has been chosen in order to see the performance of \ac{ris} under both the existence of a sensing target and a user. The \ac{ris} is modeled as a passive reflective array whose position is denoted by $(x_r,y_r )$. To explicitly model the trade-off between sensing and communication tasks, an \ac{isac} weighting parameter $\alpha \in [0,1]$ is introduced. This parameter determines how \ac{ris} reflection gain is distributed between communication enhancement toward Bob and sensing illumination toward the target.

\subsection{Channel Model}
The wireless channels between the nodes are modeled using a Rician fading model with both \ac{los} and \ac{nlos} components. The base-band channel coefficient between nodes $i$ and $j$ is expressed as

\begin{equation}
h_{ij}=h_{ij}^{\mathrm{LoS}}+h_{ij}^{\mathrm{NLoS}},
\end{equation}

The large-scale propagation loss is described by the path loss model

\begin{equation}
PL(d)=PL_{1m}+10nlog_{10}(d)+X_\sigma,
\end{equation}

where $d$ denotes the distance between nodes, $n$ represents the path loss exponent, and $X_\sigma$ is the shadow fading term modeled as a Gaussian random variable.

Small-scale fading is modeled using a clustered Rician multi-path channel. Each channel realization consists of $L=6$ taps, where the amplitude of each tap decays exponentially according to a path decay factor. This clustered model captures the multi-path characteristics typically observed in wireless channels.
To represent time-varying wireless environments, the temporal correlation between consecutive frames is modeled using an \ac{ar} process

\begin{equation}
h_t=\rho h_{t-1}+\sqrt{1-\rho^2 } w_t,
\end{equation}

where $\rho=0.92$ represents the frame correlation coefficient and $w_t$ is a Gaussian innovation term. Each candidate \ac{ris} configuration is evaluated across $T=40$ frames to capture temporal channel variations and their impact on key generation reliability.

\paragraph*{RIS Reflection Model}
The \ac{ris} introduces an additional cascaded channel through reflection. The effective \ac{ris}-assisted channel between BS and Bob can be written as

\begin{equation}
h_{R}=h_{SR}h_{RB},
\end{equation}

where $h_{SR}$ and $h_{RB}$ represent the channels between BS–\ac{ris} and \ac{ris}–Bob, respectively. 
The \ac{ris} reflection gain depends on the number of elements and the geometric alignment between incoming and outgoing signals. The effective reflection gain is approximated by

\begin{equation}
G=N\eta_{elem}(c_{in}c_{out})^{\gamma},
\end{equation}

where $N$ denotes the number of \ac{ris} elements, $\eta_{elem}$ is the reflection efficiency, $c_{in}$ and $c_{out}$ represent orientation alignment factors, and $\gamma$ is the orientation loss exponent that models the angular selectivity of the \ac{ris}.

\paragraph*{RIS Configuration and ISAC Coupling}
A five-dimensional parameter vector represents the \ac{ris} configuration

\begin{equation}
    x={\{x_r,y_r,\theta_r,N,\alpha\}}.
    \label{eq:ris_parameters}
\end{equation}

Here, $(x_r,y_r )$ denotes the \ac{ris} deployment location, $\theta_r$ represents the \ac{ris} orientation, and $N$ is the number of reflecting elements. The parameter $\alpha$ acts as an \ac{isac} coupling factor that controls how \ac{ris} resources are allocated between communication and sensing tasks.
Specifically, \ac{ris}-assisted communication gain toward Bob is scaled by $\eta_B=\alpha$, while the sensing illumination toward the target is scaled by $\eta_T=1-\alpha$. This mechanism allows the \ac{ris} configuration to balance communication enhancement and sensing accuracy dynamically.

\subsection{Performance Metrics}
This work evaluates three main performance objectives together with complementary performance indicators. Communication quality is characterized by the received \ac{snr} at Bob. Security performance is quantified by the security gap, defined as the difference between the \acp{snr} at Bob and Eve, while sensing performance is measured through the sensing gain achieved by the \ac{ris}-assisted path compared to the direct-only path.

\paragraph*{Communication Quality}
The received \ac{snr} at Bob serves as a communication metric in the considered system. It captures the combined contribution of the direct BS–Bob link and the \ac{ris}-assisted reflected path. The resulting \ac{snr} can be expressed as

\begin{equation}
    \mathrm{SNR}_B=\frac{P_t|h_{SB}+\alpha h_{SR}h_{RB}|^2}{N_0},
\end{equation}

where $P_t$ denotes the transmit power and $N_0$ is the noise power. 
Similarly, the \ac{snr} received at the target can be expressed as

\begin{equation}
    \mathrm{SNR}_T=\frac{P_t|h_{ST}+\alpha h_{SR}h_{RT}|^2}{N_0},
\end{equation}

\paragraph*{Security Performance}
To characterize the security advantage of the legitimate link over a potential eavesdropper, the security gap is defined as the difference between the average received \ac{snr} at Bob and Eve:

\begin{equation}
    \mathrm{G}_{\mathrm{sec}} = \mathrm{SNR}_B - \mathrm{SNR}_E.
\end{equation}

A larger security gap indicates a stronger separation between the legitimate and eavesdropping channels, which is beneficial for both secure communication and \ac{plkg}.

\paragraph*{Sensing Performance}
Instead of directly optimizing a localization bound, sensing performance is quantified through the sensing gain introduced by the \ac{ris}. Specifically, the sensing gain is defined as the improvement of the target-side \ac{snr} compared to the direct-only path:

\begin{equation}
    G_{\mathrm{sens}} = \mathrm{SNR}_T^{\mathrm{(total)}} - \mathrm{SNR}_T^{\mathrm{(direct)}}.
\end{equation}

This metric reflects how effectively the \ac{ris} enhances target illumination and improves sensing capability.

\subsection{Objective Formulation}
The deployment and configuration problem is formulated as a multi-objective optimization task. The aim is to determine the optimal \ac{ris} position, orientation, surface size, and \ac{isac} configuration parameter that i) achieve a suitable communication performance and sensing accuracy, and ii) incorporating security reliability into the deployment and configuration design.

Three objectives are considered in the optimization problem. The first objective is to maximize the communication quality at Bob, characterized as $\mathrm{SNR}_B$. The second objective is to maximize the security advantage of the legitimate link over a potential eavesdropper, quantified by the security gap $\mathrm{G}_\mathrm{sec}$. The third objective is to enhance sensing performance by maximizing the sensing gain of the \ac{ris}-assisted path relative to the direct-only sensing path.
The multi-objective optimization problem can therefore be expressed as

\begin{equation}
    \mathcal{N}(-\mathrm{SNR}_B(x))+\mathcal{N}(-\mathrm{G_{sec}}(x))+\mathcal{N}(-\mathrm{G_{sens}}(x)),
    \label{eq:objective}
\end{equation}

with 
\begin{equation}
    \mathcal{N}(y)=\min\left(\frac{y-\min(y)}{\max(y)-\min(y)}\right).
\end{equation}

Due to the nonlinear wireless channels and the mixture of discrete and continuous variables, the optimization problem is highly non-convex. Consequently, an iteratively sorting algorithm is adopted to approximate joint-optimal solutions. Within the test area, this iterative search algorithm repeatedly evaluates different \ac{ris} deployment coordinates. At each candidate position, it assesses the impact on communication with Bob, key consistency, and target sensing. The best-performing locations are then retained, and the search is progressively refined in their vicinity, eventually converging to an optimal deployment region.

\section{Result Analysis and Discussion} 
\label{result}
This section presents and discusses the simulation results. It first analyzes the individual objectives, followed by a heatmap illustrating the optimal \ac{ris} placement. Subsequently, representative solutions obtained by the algorithm are examined. Finally, some additional remarks on the results are provided.

\begin{figure*}[ht]
\centering
\begin{subfigure}{0.3\textwidth}
\centering
\includegraphics[width=\linewidth, height=4cm]{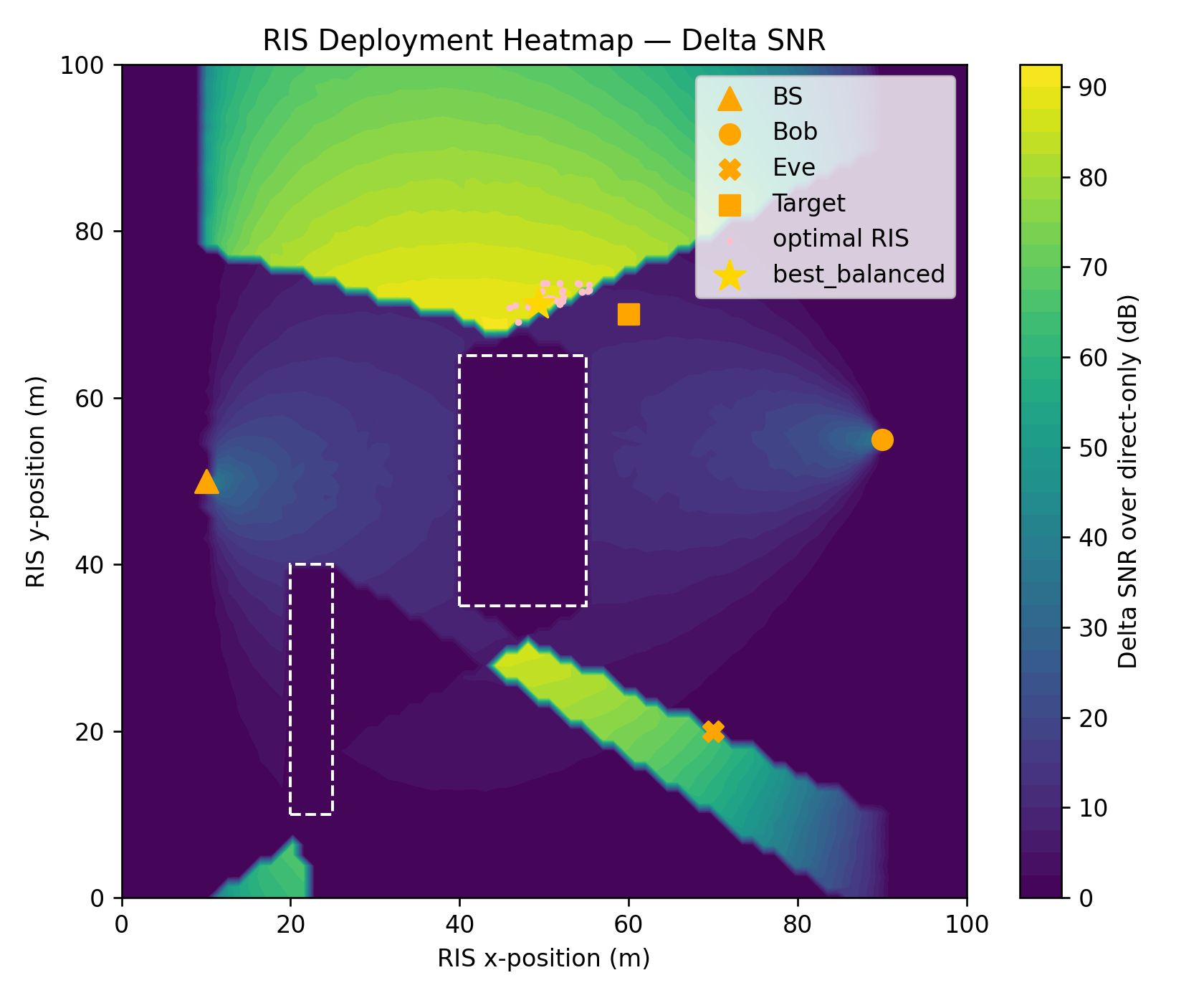} 
\caption{\ac{snr} at Bob}
\label{fig:delta_snr}
\end{subfigure}
\vspace{10pt}
\begin{subfigure}{0.3\textwidth}
\centering
\includegraphics[width=\linewidth, height=4cm]{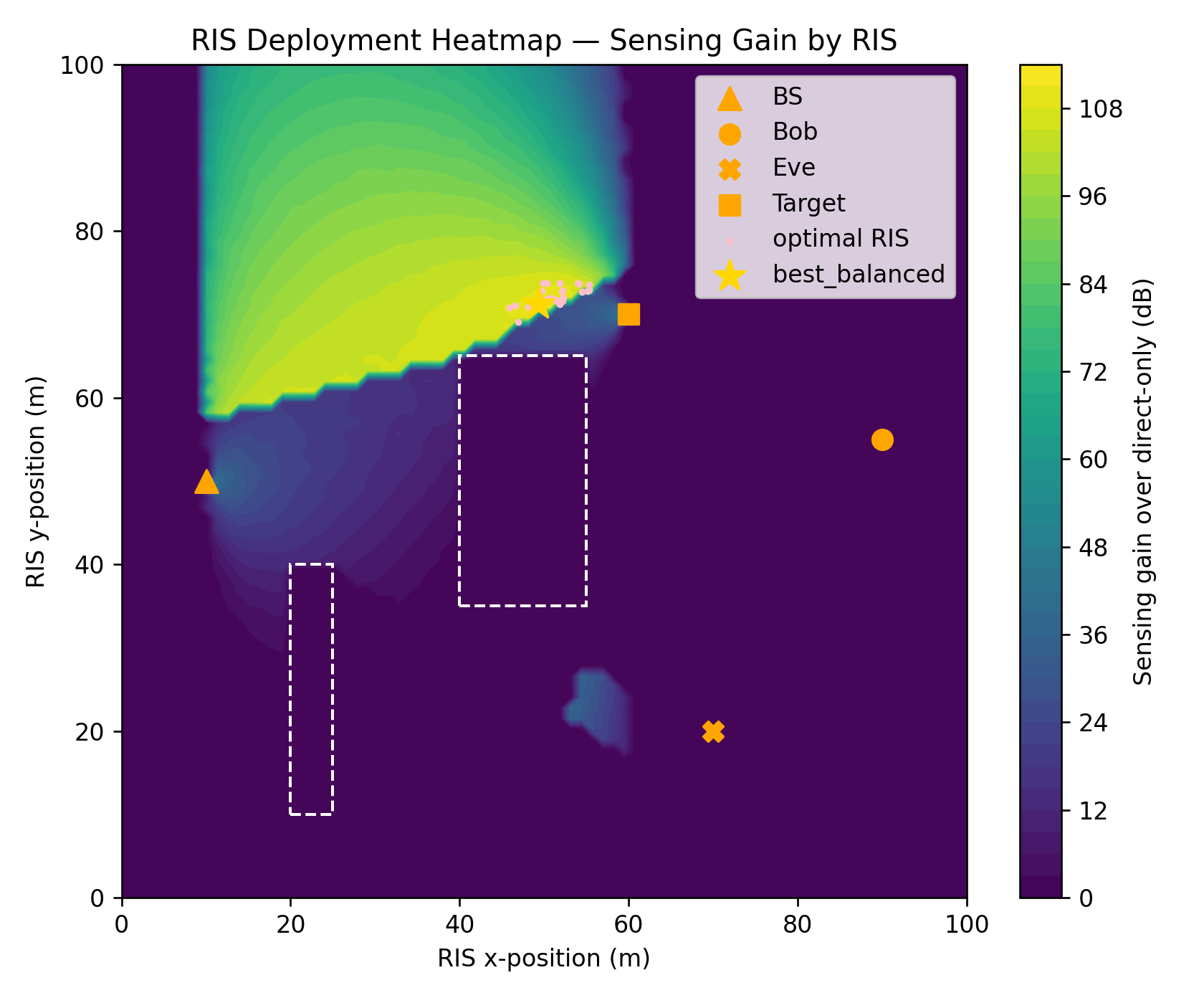}
\caption{Sensing gain}
\label{fig:sensing_gain}
\end{subfigure}
\begin{subfigure}{0.3\textwidth}
\centering
\includegraphics[width=\linewidth, height=4cm]{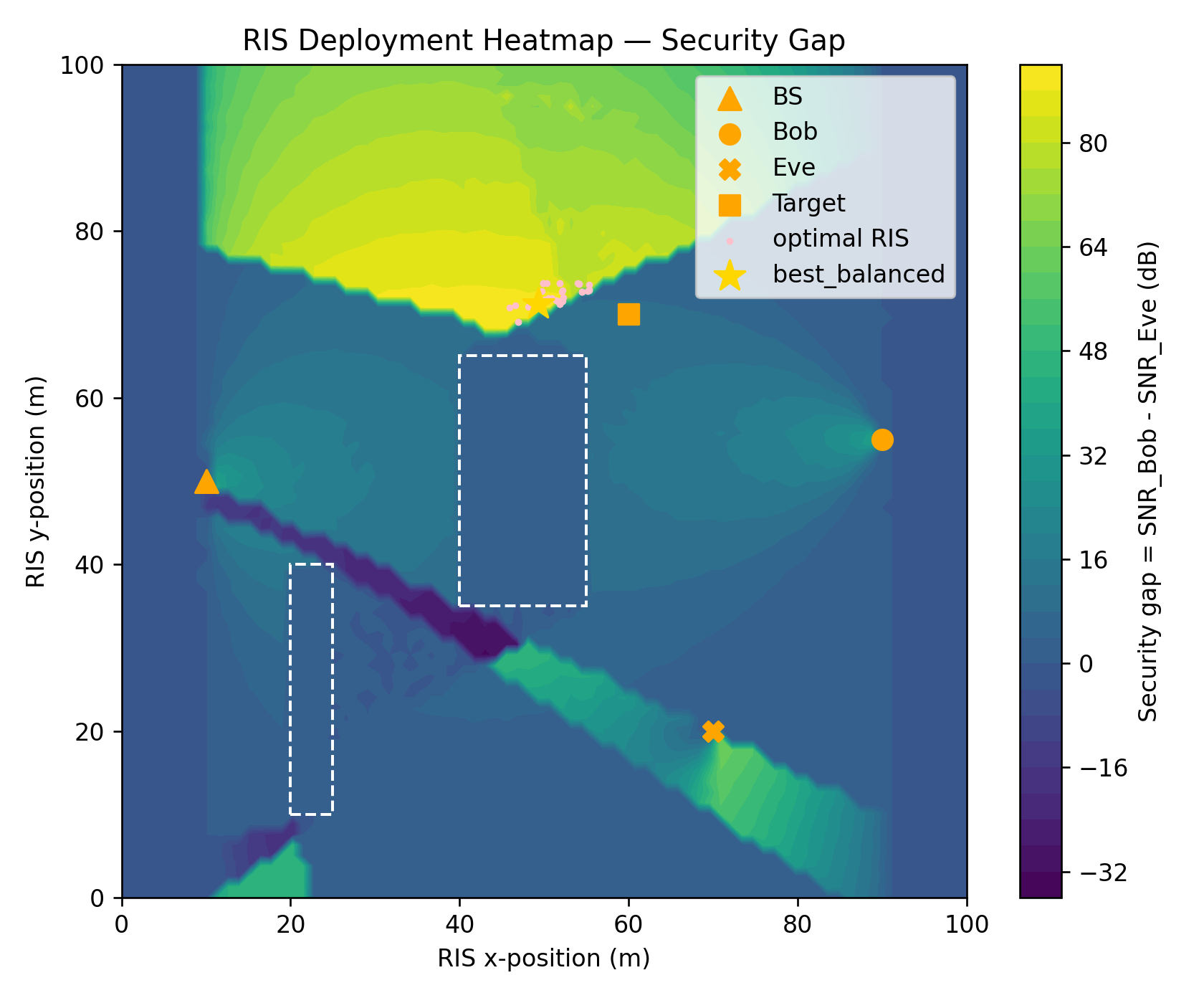}
\caption{Security gap}
\label{fig:secure_gap}
\end{subfigure}
\caption{Single objective analysis of \ac{ris} deployment: (a) \ac{snr} received at Bob over the direct link, (b) Sensing gain by the \ac{ris} over the direct link, (c) Security gap between the legitimate and eavesdropper channel}
\label{fig:performance_tradeoffs}
\end{figure*}

\subsection{Single Objective Optimization Analysis}
To visualize the spatial impact of \ac{ris} deployment on communication, security, and sensing performance,  \autoref{fig:performance_tradeoffs} presents a set of heatmaps over the considered deployment area. Each point in the map corresponds to a candidate location, while the orientation and other parameters are fixed to representative values. 

The figures illustrate the spatial distribution of the \ac{snr} improvement over the direct-only link, the sensing gain achieved by the \ac{ris}-assisted path, and the security gap between Bob and Eve. The locations of the BS, Bob, Eve, and the sensing target are also indicated. In addition, optimal solutions obtained from the multi-objective optimization are overlaid, together with a representative balanced solution.

In these three heatmaps, it can be observed that the highlighted areas are primarily concentrated in the upper region of the scene, where there are more favourable geometric paths that align with the positions of the BS and Bob, while also suppressing the link on the Eve side. Furthermore, in the sensing gain heatmap, deploying \ac{ris} in the top-left section of the map better accommodates the BS-\ac{ris}-target geometric structure, thereby providing improved sensing performance for the target.
In the lower region of the map, there are also some positions that can provide better geographical alignment for both Bob's communication performance and target sensing performance; however, as these are closer to Eve, information leakage on the Eve side is thus greater, resulting in suboptimal security performance.

\subsection{Combined RIS Deployment Analysis}

\begin{figure}[t]
    \centering
    \includegraphics[width=0.5\textwidth,height=7cm]{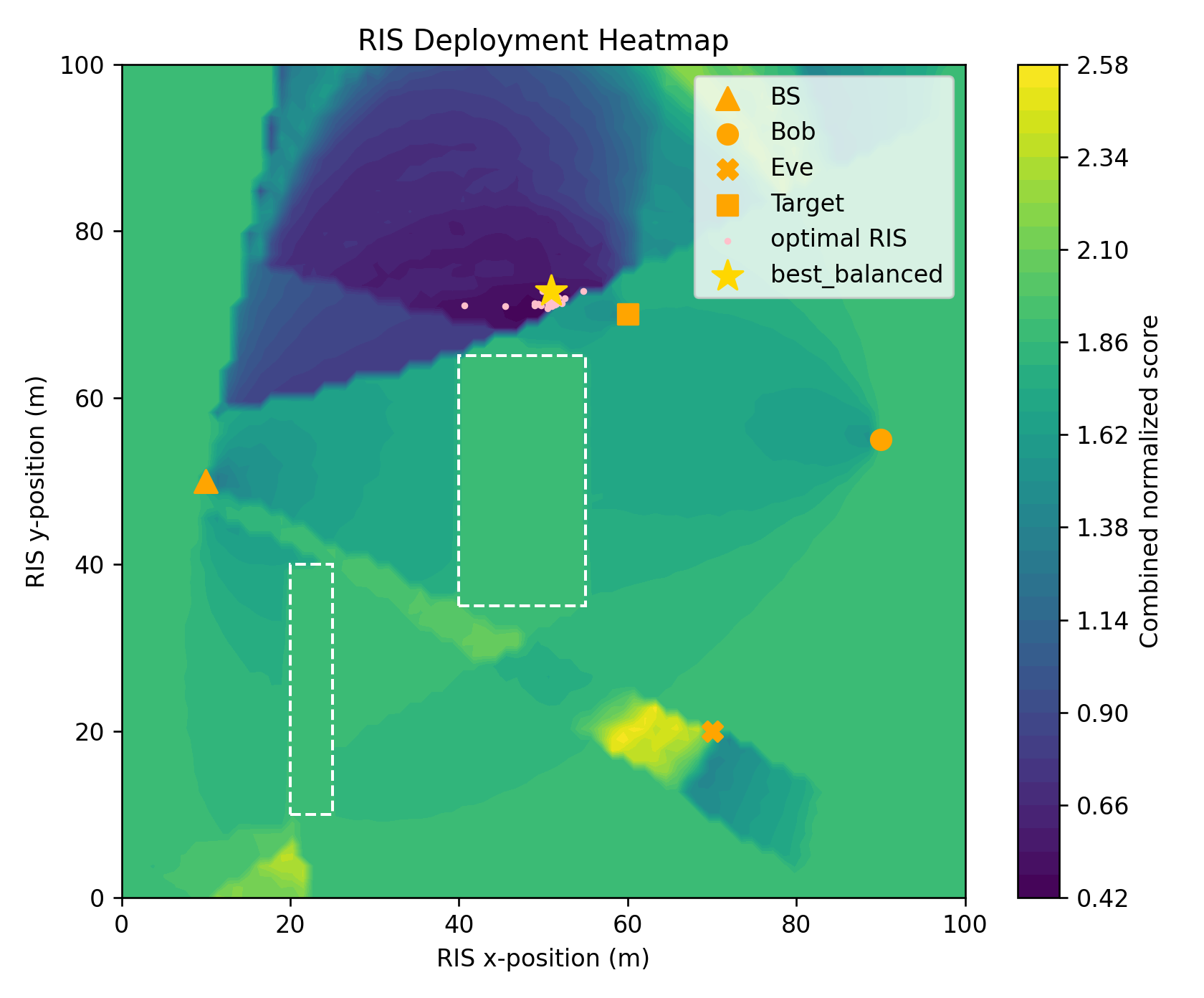}
    \caption{RIS deployment heatmap}
    \label{fig:heatmap} 
\end{figure}

The heatmap in \autoref{fig:heatmap} illustrates the spatial distribution of the optimization solution of the problem defined in \autoref{eq:objective} as a function of the \ac{ris} deployment location. Other parameters such as \ac{ris} orientation and size, as well as \ac{isac} coupling factor, are taken from the balanced solution set. The color scale visualizes the normalized optimization objective based on \ac{snr}, security gap, and sensing gain by \ac{ris}. Consistent with the multi-objective defined in \autoref{eq:objective}, lower values correspond to more favorable deployment locations, as they reflect reliable communication, accurate sensing, and reduced information inconsistencies. 

It can further be seen that the set of optimal deployment locations obtained from the five-dimensional iterative optimization cluster within the darker regions of the heatmap. This close correspondence indicates strong agreement between the identified optima and the low-value areas of the objective landscape.

This observation indicates that the deployment location is the dominant variable in the considered optimization problem. Although the remaining parameters are optimized jointly with the location, they primarily refine the solution locally, whereas the overall preference for certain spatial regions remains largely unaffected.

\subsection{Representative solutions}

\begin{table*}[t]
  \centering
  \caption{Representative results}
  \label{tab:representative_result}
  \renewcommand{\arraystretch}{1.2}
  \begin{tabular}{lcccccccc}
  \hline
Solutions & \ac{ris} position & $\theta/\,rad$ & N & $\alpha$ & $\mathrm{SNR}_B$ & $\mathrm{SNR}_T$ & security gap \\
  \hline
  Best $\mathrm{{\Delta SNR}_B}$ & (47.6,71.1) & $0.092$ & 512 & 0.95 & 62.84 & 44.61 & 61.79  \\
  Best security gap & (50.6,72.0) & $0.096$ & 512 & 0.99 & 64.0 & 21.67 & 62.39  \\
  Best sensing gain & (50.6,72.1) & $0.074$ & 512 & 0.19 & 47.41 & 73.89 & 48.21 \\
  Balanced & (50.6,72.1) & $0.074$ & 512 & 0.19 & 47.41 & 73.89 & 48.21 \\
  \hline
  \end{tabular}
\end{table*}

Table \ref{tab:representative_result} summarizes several representative \ac{ris} configurations selected from the optimal solution set, including the solutions that maximize communication performance, security, and sensing gain, as well as a balanced solution.

It can be observed that the solution achieving the highest $\mathrm{SNR}_B$ provides excellent communication performance, but does not yield the best sensing gain. In contrast, the solution that maximizes sensing gain significantly improves $\mathrm{SNR}_T$, while sacrificing communication quality. 

Interestingly, the solution that maximizes the secure gap achieves the largest separation between Bob and Eve, but does not necessarily correspond to the highest $\mathrm{SNR}_B$, indicating that security depends not only on the strength of the legitimate link but also on the relative degradation of the eavesdropping link.

The balanced solution provides a compromise among the three objectives, achieving moderate communication performance, strong sensing gain, and a relatively large secure gap. Notably, this solution shares similar configuration parameters with the sensing-optimal solution, suggesting that sensing-oriented deployments can also offer favorable security properties under certain geometric conditions.

\subsection{Remarks on the results}
In the results presented, the scaling factor is positive. This arises from the definition of $\Delta \mathrm{SNR}$, which is computed as the difference between the signal strength at the user and that of the communication link under obstacle-induced blockage. Given that the obstruction significantly attenuates the signal, the blocked link exhibits considerably lower signal strength, thereby resulting in a positive value. Likewise, the sensing gain aided by \ac{ris} at the sensing target and the security gap between Bob's and Eve's ends are calculated similarly. 

A further limitation of the current study is that obstacle-induced reflections are not considered in the signal propagation model. Incorporating such effects will be addressed in future work.

%
%
%
%
\section{Conclusion and Outlook}
\label{conclusion}

This paper investigates the joint optimization of communication, sensing, and security in \ac{ris}-assisted \ac{isac} networks. A multi-objective optimization framework was developed to determine the optimal \ac{ris} deployment location. The proposed framework jointly considers communication performance, the security gap between the legitimate user and eavesdropper, and sensing localization accuracy within a unified system perspective. To ensure realistic performance evaluation, a wireless channel model incorporating multi-path fading, blockage effects, and temporal correlation was adopted. 

The simulation results reveal inherent trade-offs among communication quality, security performance, and sensing accuracy. Specifically, configurations that prioritize sensing tend to significantly improve localization accuracy, but may simultaneously degrade communication security. In contrast, communication-oriented configurations provide higher \ac{snr} gains for legitimate users, while potentially limiting sensing performance.

The proposed framework offers a systematic approach for exploring these trade-offs and provides practical insights into efficient \ac{ris} deployment and configuration strategies for future beyond-5G and 6G wireless networks. In future work, more complex scenarios will be considered, such as multiple \acp{ris} or mobile users and targets. Future work may extend this study by considering multiple \ac{ris} deployments, mobile users and targets, and more advanced beamforming strategies such as adaptive phase control and learning-based configuration.

%
%
%
%
\section*{Acknowledgment}

This work has been supported by the Federal Ministry of Research, Technology and Space of the Federal Republic of Germany (F\"{o}rderkennzeichen 16KIS2342, IGEL-AI). The authors alone are responsible for the content of the paper.

%
%
%
%

\balance
\printbibliography

\end{document}